\documentclass[a4paper,11pt]{article}
\usepackage{pos}

\usepackage{float}
\usepackage{graphicx}% Include figure files
\usepackage{dcolumn}% Align table columns on decimal point
\usepackage{bm}% bold math
\usepackage{xspace}
\usepackage{lineno}
\usepackage{xcolor}

\newcommand{\ipp}{\ensuremath{I_{\rm pp}}\xspace}

\newcommand{\pt}{\ensuremath{p_{\rm T}}\xspace}

\newcommand{\rt}{\ensuremath{R_{\rm T}}\xspace}
\newcommand{\rtmin}{\ensuremath{R_{\rm T}^{\rm min}}\xspace}
\newcommand{\rtmax}{\ensuremath{R_{\rm T}^{\rm max}}\xspace}
\newcommand{\nmpi}{\ensuremath{N_{\rm mpi}}\xspace}

\newcommand{\py}{PYTHIA\xspace}
\newcommand{\hw}{HERWIG\xspace}

\usepackage{bibspacing}
\title{Particle production as a function of the underlying event in pp collisions simulated with PYTHIA 8}
\ShortTitle{Particle production based on the underlying event in pp collisions with PYTHIA 8}

\author*[a]{Gyula Benc\'edi}
\emailAdd{gyula.bencedi@cern.ch}
\author[a]{Antonio Ortiz}
\affiliation[a]{%
Instituto de Ciencias Nucleares, Universidad Nacional Aut\'onoma de M\'exico,\\
 Apartado Postal 70-543, M\'exico Distrito Federal 04510, M\'exico %\textbackslash\textbackslash
}
\author[b]{Antonio Paz}
\affiliation[b]{%
Facultad de Ciencias F\'isico Matem\'aticas,
Universidad Aut\'onoma de Nuevo Le\'on,\\ 
Ciudad Universitaria, San Nicolas de los Garza, Nuevo Le\'on 66450, M\'exico
}

\abstract{
In this work we present the production of charged particles associated with high-$p_{\rm T}$ trigger particles ($8<p_{\rm T}^{\rm trig.}<15$ GeV /$c$) at mid-pseudorapidity in proton-proton collisions at $\sqrt{s}=5.02$\,TeV simulated with the \textsc{PYTHIA 8} Monte Carlo model. The study is performed as a function of the relative transverse activity classifier, $R_{\rm T}$, which is the relative charged-particle multiplicity in the transverse region ($\pi/3 <|\Delta\phi| <2\pi/3$) of the di-hadron correlations, and it is sensitive to the Multi-Parton Interactions. The evolution of the yield of associated particles on both the toward and the away regions with $3\leq p_{\rm T}^{\rm assoc.}<8$ GeV/$c$ as a function of $R_{\rm T}$ is investigated. We propose a strategy which allows for the modelling and subtraction of the Underlying Event (UE) contribution from the toward and the away regions in challenging environments like those characterised by large $R_{\rm T}$. We found that the signal in the away region becomes broader with increasing $R_{\rm T}$. Contrarily, the yield increases with $R_{\rm T}$ in the toward region. This effect is reminiscent of that seen in heavy-ion collisions, where an enhancement of the yield in the toward region for $0-5\%$ central Pb--Pb collisions at $\sqrt{s}_{\rm NN}=2.76$\,TeV was reported. To further understand the role of the UE and additional jet activity, the transverse region is divided into two one-sided sectors, ``trans-max'' and ``trans-min'' selected in each event according to which region has larger or smaller charged particle multiplicity. Based on this selection criterion, the observables are studied as a function of $R_{\rm T}^{\rm max}$ and $R_{\rm T}^{\rm min}$, respectively. The presented results have been published in Refs.~\cite{Bencedi:2020qpi,Bencedi:2021tst}. Reference~\cite{Bencedi:2021tst} contains detailed studies with the \textsc{HERWIG 7.2} Monte Carlo generator as well.
}

\FullConference{%
 The Ninth Annual Conference on Large Hadron Collider Physics - LHCP2021\\
 7-12 June 2021\\
 Online
}

\begin{document}
\maketitle

\section{Introduction}

One of the  most  important  discoveries at the LHC is the presence of collectivity~\cite{Nagle:2018nvi} in high-multiplicity proton-proton (pp) collisions which is originally present in heavy-ion collisions (AA) where the effect of jet quenching is observed~\cite{Wiedemann:2009sh}. To date no jet quenching is found in small collision systems, however several attempts have already been made to unveil it in high-multiplicity pp collisions, see e.g. Ref.~\cite{Jacobs:2020ptj}. 
To extract jet-quenching signal from pp data, one has to deal with the selection bias which is the main technical issue in the analysis~\cite{Jacobs:2020ptj}. The high-activity event selection results a sample which is naturally biased toward hard processes (including jets).
% To overcome this difficulty we investigate the particle production as a function of a new variable \rt introduced recently, and discuss a strategy to minimize such biases in order to have an observable more suitable for jet quenching searches in small systems.
To overcome this difficulty, it has been proposed explicitly removing the jet contribution from the event activity estimator. This can be achieved if one identifies an axis which allows for the event-by-event separation of the jet contribution from the underlying event (UE). The UE consists of particles that arise from beam-beam remnants and Multiparton Interactions (MPI), the latter being originated from soft and semi-hard scatterings~\cite{Field:2012kd}. The direction of the leading charged-particle transverse momentum (\pt) can be used as a reference axis to build particle correlations with the associated particles in azimuthal angle $\Delta\phi$. Then, the particle production can be studied as a function of the charged-particle multiplicity in the following geometrical regions: toward region ($|\Delta\phi|<\pi/3$), away region ($|\Delta\phi|>2\pi/3$), and transverse region ($\pi/3 <|\Delta\phi| <2\pi/3$). The toward and the away regions are dominated by string fragments originating from the hardest partonic process of the event, and are expected to be nearly insensitive to the softer UE. In contrast, the trans. region is the most sensitive to UE, but it has contributions from initial- and final-state radiation (ISR and FSR) that accompanies the hard scattering. This UE region will be used to isolate events with exceptionally large or small activity with respect to the event-averaged mean.

In this work we explore a quantity called \ipp which is motivated by the $I_{\rm AA}$~\cite{STAR:2006vcp,PHENIX:2010sgo,ALICE:2011gpa} commonly used in AA collisions to study jet quenching effects. $I_{\rm AA}$ is the ratio of jet-like yield from AA to the one from pp collisions. In the toward region, the $I_{\rm AA}$ provides information about the fragmenting jet leaving the medium, while in the away region it additionally reflects the probability that the recoiling parton survives the passage through 
the medium. For central Pb--Pb collisions, a significant suppression ($I_{\rm AA}\approx0.6$) in the away region is reported, whereas only a moderate enhancement ($I_{\rm AA}\approx1.2$) was observed in the toward region~\cite{ALICE:2011gpa,CMS:2012uxa}; these effects indicate the presence of the QGP. We further investigate the UE-dominated trans. region considering a refinement to distinguish between the more and the less active sides of the trans. region on a per-event basis~\cite{Marchesini:1988hj,Pumplin:1997ix,CDF:2004jod}. The purpose of using this geometrical selection criterion is to suppress hard ISR/FSR thus increasing the sensitivity of the trans. region to MPI component of the UE. Based on the CDF approach~\cite{Field:2012kd},  the so-called trans-min and trans-max regions are introduced. The trans-min region is insensitive to wide angle emissions from the hard process, while the trans-max region receives contribution from hard ISR and/or FSR. We study the structures of the $\Delta\phi$ distribution for the trans-max and trans-min regions. The \pt distributions of charged particles, as well as the proton-to-pion ratio, are also studied as a function of multiplicity in the trans-min and trans-max regions. The results are discussed in the context of recent ALICE preliminary results~\cite{Tripathy:2021fax,Nassirpour:2020owz}.

\section{Classification of event activity using the underlying event}

We simulated inelastic pp collisions at $\sqrt{s}=5.02$\,TeV and only primary charged particles~\cite{ALICE-PUBLIC-2017-005} with $\pt>0.5$\,GeV/$c$ and pseudorapidity $|\eta|<0.8$ are considered in the analysis. 
% For $\pt^{\rm trig.}>8-10$\,GeV/$c$, the mean charged-particle multiplicity in the transverse region, $\langle N_{\rm ch}^{\rm trans.} \rangle$, has only a weak dependence on $\pt^{\rm trig.}$~\cite{Acharya:2019nqn}. Therefore, we focus on events having $\pt^{\rm trig.}$ above the onset of the plateau, and we classify events with a trigger particle in the range $8\leq\pt^{\rm trig.}<15$\,GeV/$c$ based on their per-event activity in the transverse region with respect to the mean:
We use the Relative Transverse Activity Classifier, \rt, which has been recently introduced~\cite{Martin:2016igp,Ortiz:2017jaz} and used on LHC data~\cite{ALICE:2019mmy,Zaccolo:2019hxt}. \rt uses the conventional definition of the transverse region, which was adopted in the UE analysis originally introduced by the CDF collaboration~\cite{Field:2000dy,CDF:2015txs}. We classify events with a trigger particle in the range $8\leq\pt^{\rm trig.}<15$\,GeV/$c$ based on their per-event activity in the trans. region with respect to the mean: $\rt=N_{\rm ch}^{\rm trans.}/\langle N_{\rm ch}^{\rm trans.} \rangle$.
% where the mean charged-particle multiplicity in the transverse region has only a weak dependence on $\pt^{\rm trig.}$~\cite{Acharya:2019nqn}.
% The study is performed as a function of the relative transverse activity classifier, $R_{\rm T}$, which is the relative charged-particle multiplicity in the transverse region ($\pi/3 <|\Delta\phi| <2\pi/3$) of the di-hadron correlations, and it is sensitive to the Multi-Parton Interactions.
Also, we define two regions that are characterized in terms of their relative charged-particle multiplicities, $N_{\rm ch}^{\rm trans.\,max}$ and $N_{\rm ch}^{\rm trans.\,min}$, termed trans-max and trans-min. Trans-max (trans-min) refers to the trans, region containing the largest (smallest) number of charged particles. Using $N_{\rm ch}^{\rm trans.\,min}$ and $N_{\rm ch}^{\rm trans.\,max}$, instead of $N_{\rm ch}^{\rm trans.}$, we also can define the quantities $R_{\rm T}^{\rm min}$ and $R_{\rm T}^{\rm max}$, respectively. For $\rt,\rtmax \approx 8$ and $\rtmin \approx 8$, the inclusive multiplicity reach is about 6 times and 4 times the mean multiplicity. The average \nmpi increases with \rtmin as opposed to what is observed for \rtmax. Results of the correlation between the average \nmpi and \rt showed that events selected with $\rt>2$ but with similar \nmpi include additional jets to enhance the activity in the trans. region which in turn causes a selection bias.

% with the same \nmpi but with different multiplicities in the transverse region suggests that at some point particles from additional jets may be picked up to enhance the activity in the transverse region. In turn, above a given \rt value, the activity in the transverse region can be biased towards harder processes due to the presence of a third jet in the transverse region~\cite{Field:2002vt}.

\section{Monte Carlo models and the extraction of jet-like signals}

For our analyses~\cite{Bencedi:2020qpi,Bencedi:2021tst}, we used \py~8.201~\cite{Sjostrand:2007gs} with the default Monash 2013 tune~\cite{Skands:2014pea} and \hw~7.2~\cite{Bellm:2019zci} with its default tune SoftTune~\cite{Bellm:2019zci}. Both models are fully exclusive hadron-level MC generators, containing leading-logarithmic initial- and final-state parton showers, hadronization models and particle decays. The focus of the mentioned tunes for these generators is on the description of minimum-bias as well as underlying-event data.

\begin{figure*}
\centering
\includegraphics[width=0.33\columnwidth, keepaspectratio]{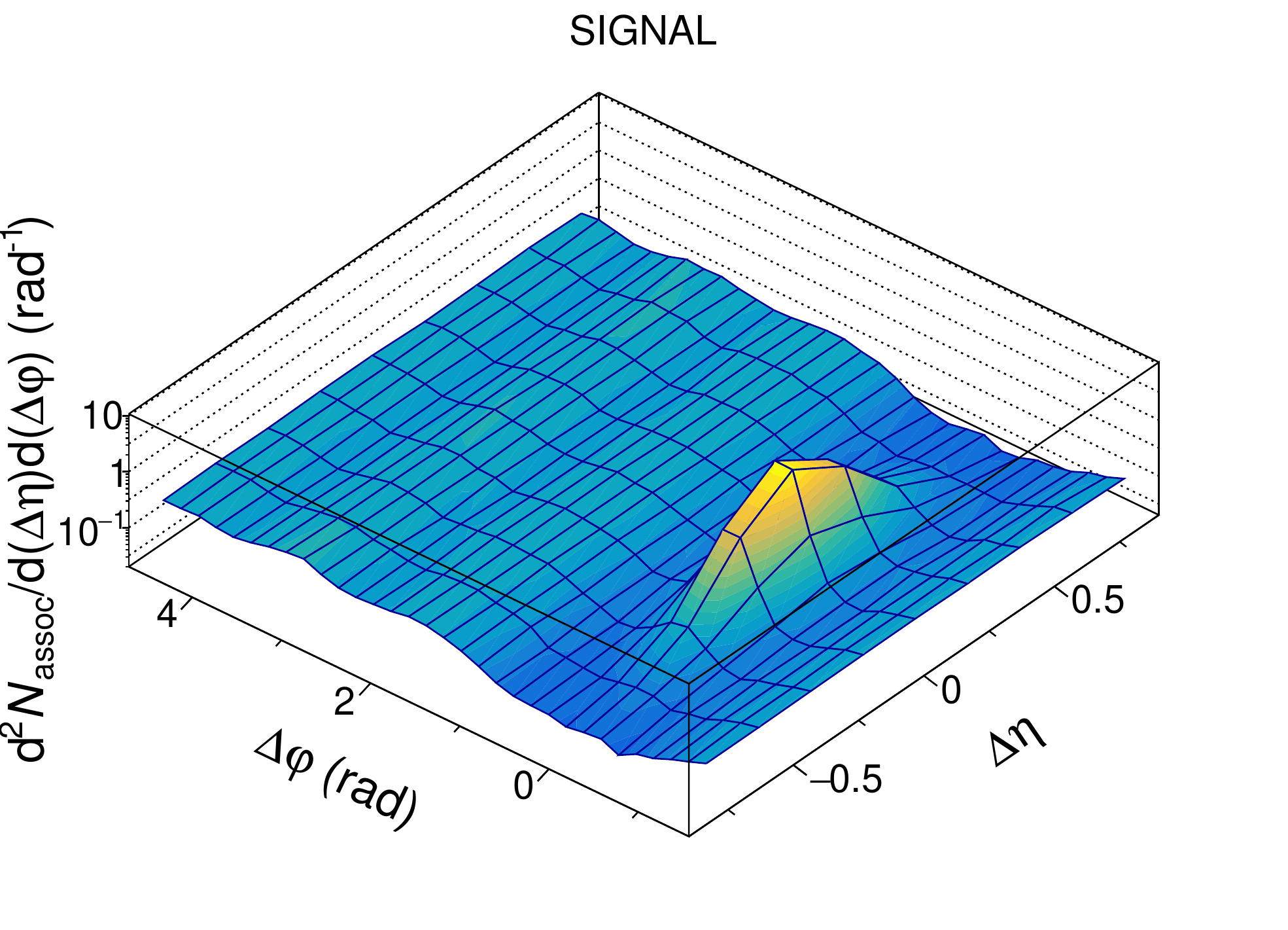}%
\includegraphics[width=0.33\columnwidth, keepaspectratio]{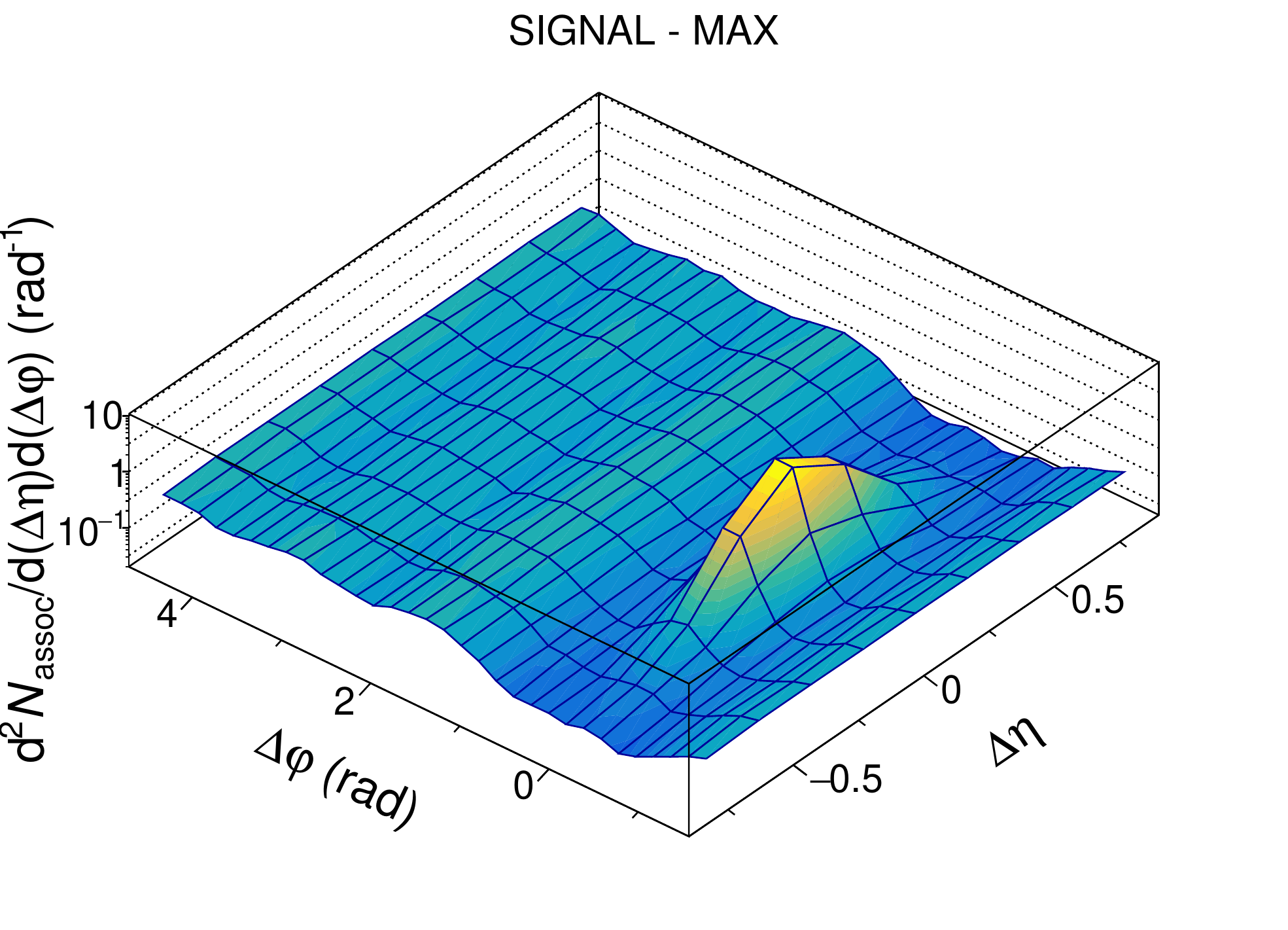}%
\includegraphics[width=0.33\columnwidth, keepaspectratio]{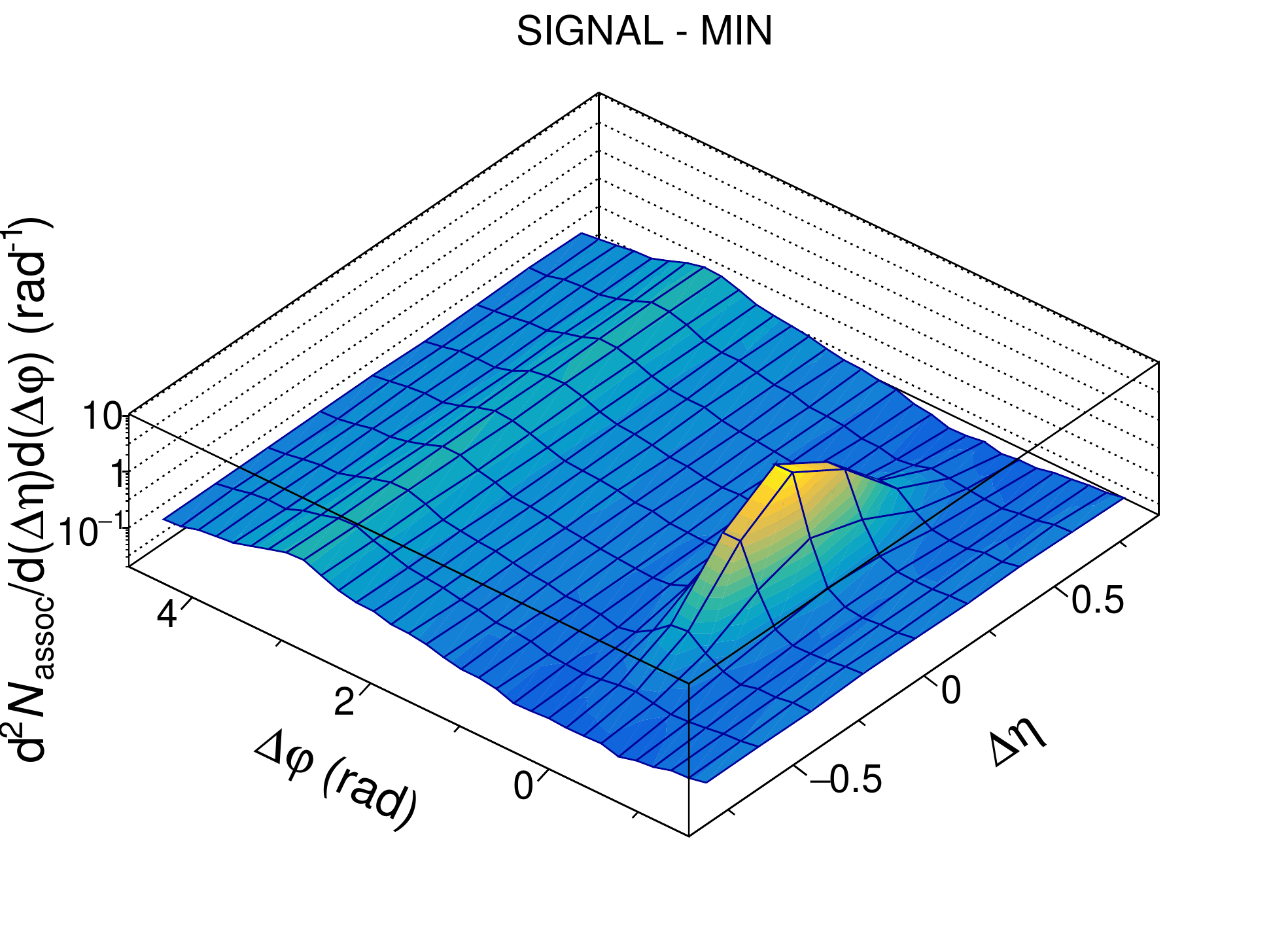}%
    \caption{\label{fig:dphideta} 
    Charged particle yield as a function of $\Delta\eta-\Delta\phi$ for $\rt\geq3.5$ and $4\leq\pt^{assoc.}<6$\,GeV/$c$ simulated with \py~8 in pp collisions at $\sqrt{s}=5.02$\,TeV. Second and third columns represent the cases of trans-max and trans-min corresponding to high and low activities in the transverse region. 
    }
\end{figure*}

Figure~\ref{fig:dphideta} shows the charged particle yield as a function of $\Delta\eta-\Delta\phi$ for $\rt\geq3.5$. A structure elongated in $\Delta\eta$ is present in the trans. region which is a consequence of the event selection. Therefore, its contribution to the toward and the away regions has to be removed. For this purpose mixed events are used to model the uncorrelated contribution as well as the acceptance effect. The signal is extracted after removing the mixed event distribution from the same event distribution. The $\Delta\eta$-$\Delta\phi$ distributions are flat in $\Delta\eta$, therefore, the underlying event contribution to the toward and away regions is subtracted with the zero yield at minimum assumption~\cite{Ajitanand:2005jj}.

\section{Results}

\begin{figure}[t]
\centering
\includegraphics[width=0.75\textwidth]{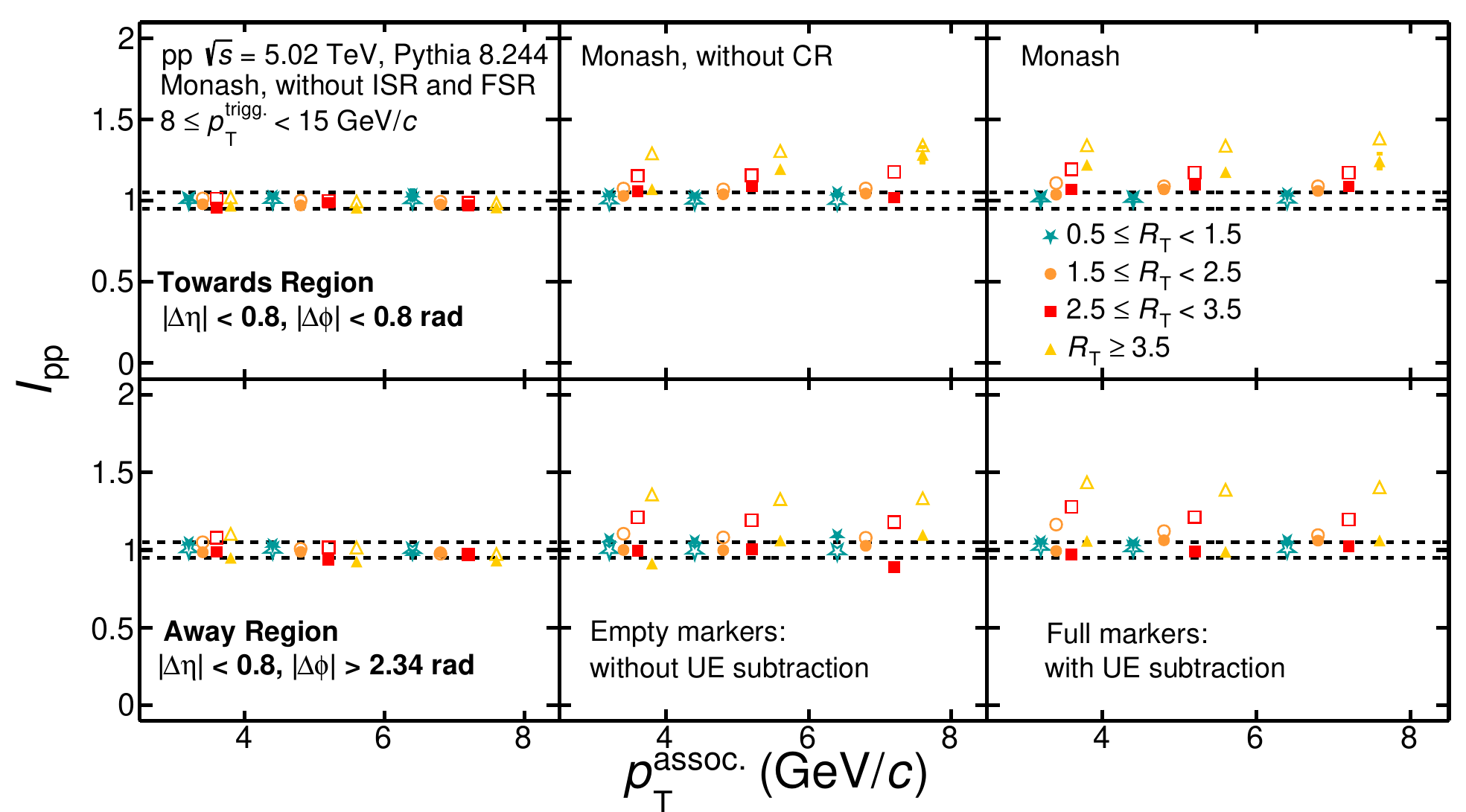}% Here is how to import EPS art
\caption{\label{fig:4} $I_{\rm pp}$, the ratio of yield from different \rt classes and from the \rt-integrated class as a function of $R_{\rm T}$ for the toward (top) and away region (bottom). 
Dashed lines indicate $\pm5\%$ deviation from unity.}
\end{figure}

We calculated the $I_{\rm pp}$ quantity, which is the ratio of yields from different \rt classes to the \rt-integrated one. In the absence of either selection bias or jet quenching, $I_{\rm pp}$ is expected to be consistent with unity. The influence on the jet-like signals of the remaining structures are reduced by integrating the $\Delta\phi$ distribution within the interval $|\Delta\phi|<0.8$ ($|\Delta\phi-\pi|<0.8$) for the toward (away) region. Figure~\ref{fig:4} shows the $I_{\rm pp}$ as a function of $\pt^{\rm assoc.}$ for both the toward and the away region. Three $\pt^{\rm assoc.}$ intervals are considered: $3\leq\pt^{\rm assoc.}<4$\,GeV/$c$, $4\leq\pt^{\rm assoc.}<6$\,GeV/$c$, and $6\leq\pt^{\rm assoc.}<8$\,GeV/$c$. Results with and without underlying event subtraction are also presented. We observe that the \ipp quantity is independent of \rt in simulations which do not include ISR and FSR. With physical settings (including ISR/FSR) the yield in the toward region increases with \rt; this effect is reminiscent of the Pb--Pb data from the ALICE and CMS experiments, although in the \py~8 model it is driven by a bias, and has to be taken into account in any analysis which use \rt as an event classifier. The bias originates from hard Bremsstrahlung gluons that can produce an apparent modification of the jet-like yield in events with extremely large underlying event. In contrast, the \ipp in the away region is consistent with unity, and independent of \rt and $\pt^{\rm assoc.}$. This suggests that \py~8 can partially mimic jet-quenching effects.

\begin{figure*}
\begin{center}
%\begin{figure}[t]
\includegraphics[width=0.96\textwidth]{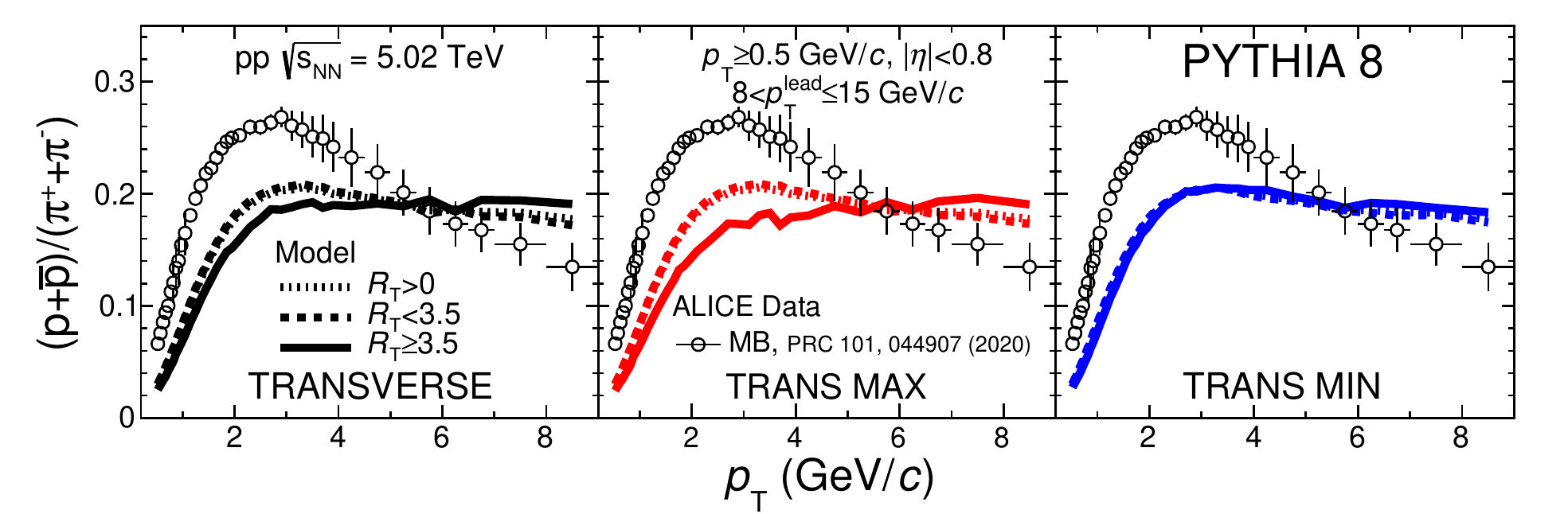}
\caption{Proton-to-pion ratio as a function of \pt for the transverse (left panel), trans-max (middle panel), and trans-min (right panel) regions for pp collisions at $\sqrt{s}=5.02$\,TeV.}
\label{fig:ptopi}  
\end{center}
\end{figure*}

We modified of the original \rt definition: split the transverse region into the trans-min and trans-max regions in order to control the sensitivity to hard processes. Using \rtmin instead of \rt or \rtmax, we do not observe a remarkable evolution of both the toward and away regions with increasing \rtmin, moreover, the remaining signal in the transverse region is much smaller and roughly \rtmin independent. 
The \pt dependent p/$\pi$ ratios shown in Fig.~\ref{fig:ptopi} for various \rt event classes exhibit a depletion at $\pt\approx2-3$\,GeV/$c$ with increasing \rt which is consistent with the presence of jets in the trans. region. As reported earlier both in MC~\cite{Ortiz:2015ttf} and data~\cite{Zimmermann:2015npa}, the ratio is smaller in the jet region than in the UE region. The \rtmax-dependent ratios receive a large amount of particles from hard Bremsstrahlung gluons produced in ISR/FSR radiation.
The middle panel of Fig.~\ref{fig:ptopi} reports the high-activity trans. region which receives a large amount of particles from hard gluon-radiated processes, i.e. from ISR/FSR. In effect, the bump structure is smeared out; a similar effect is seen as for particle ratios inside jets~\cite{Zimmermann:2015npa}. In contrast, the \rtmin-dependent ratios shown in the right panel of Fig.~\ref{fig:ptopi} exhibit an opposite behaviour: only a weak dependence on the \rtmin event classes is seen. However, the position of its maximum is slightly shifted to higher \pt values in events with large \rtmin, which is expected in \py~8 events with large \nmpi and color reconnection~\cite{OrtizVelasquez:2013ofg,Ortiz:2020rwg,Ortiz:2021peu}.

% The peak in the away region exhibits a broadening which becomes more evident for high RT values. The Ipp in the away region, on the other hand, is found to be consistent with unity, and independent of RT and passoc.T . This suggests that PYTHIA 8 with ISR and FSR can partially mimic jet-quenching effects in small systems.
% 
% Contrary to the expectations, the preliminary results of the ALICE Collaboration~\cite{} indicate that the proton-to-pion ratio does not exhibit the characteristic enhancement at intermediate \pt in events with large \rt with respect to minimum-bias pp collisions.
% The effect is a consequence of a selection bias attributed to wide-angle gluon emissions which creates jets that populate the transverse region. A modified version of \rt is used in order to suppress its sensitivity to hard gluon Bremsstrahlung, and enhance the sensitivity to soft MPI.

\section{Summary}

We study the charged particle production as a function of the underlying event. The study has been conducted using pp collisions at $\sqrt{s}=5.02$\,TeV simulated with PYTHIA 8. The goal of the study is to show that using the multiplicity in the transverse region of the di-hadron correlations, then one enhances the sensitivity to Multiparton Interactions. We propose a refinement of the event classifier by means of extracting the multiplicity in the transverse min region aiming at reducing  the remaining contributions from ISR and FSR from the main partonic scattering. The impact on observables like di-hadron correlations and particle ratios is discussed. 

The presented results encourage performing a jet quenching search using \rtmin as event activity estimator which significantly reduces the selection biases.

%  We proposed a modification of the original \rt definition: split the transverse region into the trans-min and trans-max regions in order to control the sensitivity to hard processes. Using \rtmin instead of \rt or \rtmax, we do not observe a remarkable evolution of both the toward and away regions with increasing \rtmin, moreover, the remaining signal in the transverse region is much smaller and roughly \rtmin independent. The results encourage performing a jet quenching search using \rtmin as event activity estimator which significantly reduces the selection biases.

% This bias originates from the interplay of two contributions. On the one hand, the transverse region (\rt) is very sensitive to ISR and FSR, therefore, in events with large \rt (large ISR and FSR) the transverse region often contains a third jet. On the other hand, it has been reported that the high-\pt trigger selection also biases the sample towards events with a large amount of gluon radiation. These effects produce a correlation between the activity in the transverse region and the leading \pt. 

% We emphasize that this selection bias has to be taken into account in any analysis which use \rt as an event classifier. At the same time, the peak in the away region exhibits a broadening which becomes more evident for high \rt values.  The $I_{\rm pp}$ in the away region, on the other hand, is found to be consistent with unity, and independent of \rt and $\pt^{\rm assoc.}$. This suggests that \py~8 with ISR and FSR can partially mimic jet-quenching effects in small systems.

\section{Acknowledgments}

This work has been supported by CONACyT under the Grants No. A1-S-22917 and CF-2042. G. B. acknowledge the postdoctoral fellowship grant provided by CONACyT under the Grant No. A1-S-22917. The support for part of this work has been received during the pandemic by the National Research, Development and Innovation Office (NRDIO) OTKA K120660 and FK131979,2019-2.1.11-TET-2019-00078(HuMex) and 2019-2.1.6-NEMZKI-2019-00011 (CERN).

% G.~B. acknowledges the postdoctoral fellowship of CONACyT under the Grant No. A1-S-22917. 
% We acknowledge the technical support of Luciano Diaz and Eduardo Murrieta for the maintenance and operation of the computing farm at ICN-UNAM. 

\end{document}